\documentclass[epj]{svjour}
\usepackage{amsfonts,amssymb,amsmath}
\usepackage{graphicx,color}
\usepackage{dcolumn,slashed}
\usepackage{adjustbox}
\usepackage{cases}

\makeatletter
\DeclareRobustCommand{\cev}[1]{
  {\mathpalette\do@cev{#1}}
}
\newcommand{\do@cev}[2]{
  \vbox{\offinterlineskip
    \sbox\z@{$\m@th#1 x$}
    \ialign{##\cr
      \hidewidth\reflectbox{$\m@th#1\vec{}\mkern4mu$}\hidewidth\cr
      \noalign{\kern-\ht\z@}
      $\m@th#1#2$\cr
    }
  }
}
\makeatother

\begin{document}

\title{Wigner SU(4) symmetry, clustering, and the spectrum of {\boldmath$^{12}$}C}

{\color{red}
\author{Shihang Shen \inst{1} 
\and Timo A. L\"{a}hde \inst{1} 
\and Dean Lee\inst{2}  
\and Ulf-G.~Mei{\ss}ner\inst{3,1,4}
%
}                     
%
%
\institute{
Institut~f\"{u}r~Kernphysik,~Institute~for~Advanced~Simulation and
J\"{u}lich~Center~for~Hadron~Physics, \\ 
Forschungszentrum~J\"{u}lich, D-52425~J\"{u}lich,~Germany
\and Facility for Rare Isotope Beams and Department of Physics and Astronomy, Michigan State University, \\ 
MI 48824, USA
\and Helmholtz-Institut~f\"{u}r~Strahlen-~und~Kernphysik~and~Bethe~Center~for
Theoretical~Physics, Universit\"{a}t~Bonn, \\ D-53115~Bonn,~Germany
\and Tbilisi State University, 0186 Tbilisi, Georgia
}
}

\date{Received: date / Revised version: date}
%

\abstract{
We present lattice calculations of the low-lying spectrum of $^{12}$C using a simple nucleon-nucleon interaction that is independent of spin and isospin and therefore invariant under Wigner's SU(4) symmetry.  We find strong signals for all excited states up to $\sim 15$~MeV above the ground state, and explore the
structure of each state using a large variety of 
$\alpha$ cluster and harmonic oscillator trial states, projected onto given irreducible representations of the
cubic group. We are able to verify earlier
findings for the $\alpha$ clustering in the Hoyle state and the second $2^+$ state of $^{12}$C. The success of these calculations to describe the full low-lying energy spectrum using spin-independent interactions suggest that either the spin-orbit interactions are somewhat weak in the $^{12}$C system, or the effects of $\alpha$ clustering are diminishing their influence.  This is in agreement with previous findings from {\it ab initio} shell model calculations.
} 

\maketitle


\section{Motivation}
In Ref.~\cite{Lu:2018bat}, it was shown that the ground state properties of many light and medium-mass nuclei as well as neutron matter could be well-described by microscopic $A$-body lattice calculations using a spin- and isospin-independent interaction plus the Coulomb interaction. These spin- and isospin-independent interactions obey Wigner's SU(4) symmetry \cite{Wigner:1937}, where the four spin and isospin nucleon components transform as an SU(4) quartet. The very simple interactions in Ref.~\cite{Lu:2018bat} have only four parameters: the strength of the two-nucleon interaction, the range of the two-nucleon interaction, the range of the local\footnote{The meaning of the term ``local'' is explained in the next section.} part of the two-nucleon interaction, and the strength of the three-nucleon interaction.  The strength and range of the SU(4)-symmetric local interaction were shown in Ref.~\cite{Elhatisari:2016owd} to play an important role in the effective interactions between two $\alpha$ clusters and the binding of nuclei with more than four nucleons.  These findings are consistent with empirical observations seen in other calculations \cite{Contessi:2017rww}, and the underlying mechanisms of the effective cluster interactions have been studied in detail \cite{Rokash:2016tqh,Kanada-Enyo:2020zzf}.  

An interesting unresolved question is whether or not such simple SU(4)-symmetric interactions can reproduce more than just the average ground state properties of atomic nuclei.  It is clear that nuclei with large spin-orbit splittings among nuclear subshells will not be properly described if the nucleonic interactions are independent of spin.  However, if we are considering a nuclear system where $\alpha$ clustering is important, then the impact of the spin-orbit interactions might be significantly reduced.  But this also raises the possibility that some of the nuclear states are well described while others are not.  Therefore, the success or failure of these simple spin-independent interactions in describing the spectrum of a given nucleus provides a useful probe for illuminating the underlying physics.  We should note that several {\it ab initio} shell model calculations have found that the spin-orbit splittings are not strong for $^{12}$C \cite{Hayes:2003ni,Johnson:2014xda}, suggesting that the simple SU(4)-symmetric interactions might capture the essential physics.

In this work, we focus on the nucleus with perhaps the most interesting and astrophysically important spectrum, $^{12}$C.  In the case of $^{12}$C, there is much evidence of competition between the arrangement of nucleons into shell model orbitals and the grouping of nucleons into $\alpha$ clusters \cite{Freer:2017gip}.  For the studies here, we use a simple SU(4)-symmetric interaction that was introduced in Ref.~\cite{Frame:2020mvv} for calculations of light nuclei and light hypernuclei.  In this greatly simplified interaction, there is no Coulomb interaction and no three-nucleon interaction.  There are only two free parameters:  the strength of the two-nucleon interaction and the range of the local part of the interaction.  As we will show, the results using such a simple interaction are surprisingly good.  While the interactions are extremely simple, we find evidence that the $^{12}$C nucleus sits at an interesting tipping point where the competition between the shell structure and clustering produces a low-energy spectrum with qualitatively different types of nuclear states.

While our results here are fully microscopic $A$-body calculations with full correlations to all orders, they are not {\it ab initio} in the sense that the nucleon-nucleon interaction is not fitted to nucleon-nucleon scattering data but rather tuned to reproduce the ground state energies of $^4$He and $^{12}$C.  Nevertheless, these calculations serve as an important comparative benchmark for future {\it ab initio} calculations using chiral effective field theory. 

\section{Background}
Wigner SU(4) symmetry is known to be connected to the large-color (large $N_c$) limit of QCD \cite{Kaplan:1995yg,Kaplan:1996rk}.
While the low-energy $S$-wave nucleon-nucleon interactions should satisfy Wigner SU(4) symmetry up to effects
of order $1/N_c^2$, the existence of a bound state in the spin-triplet channel and not the spin-single channel appears to violate
this result. Recently, it has been shown that the spin-isospin exchange symmetry is recovered to
the expected accuracy, provided that the momentum resolution scale (or EFT cutoff) is chosen appropriately,
at $\sim 500$~MeV~\cite{Lee:2020esp}, in agreement with earlier studies~\cite{Timoteo:2011tt}. This also provides justification for the use of a Wigner SU(4) symmetric leading-order (LO) interaction~\cite{Lu:2018bat} as a basis for the chiral EFT expansion.

Nuclear Lattice Effective Field Theory (NLEFT) is an \textit{ab initio} method where the chiral EFT
expansion is combined with Projection Monte Carlo (PMC) simulations~\cite{Lee:2008fa,Lahde:2019npb}.
For PMC, Wigner SU(4) symmetric interactions afford significant computational advantages, due to their
simplicity and minimal fermion sign oscillations.
Remarkably, recent NLEFT studies have found that the interactions between $\alpha$ particles (or $^{4}$He nuclei)
appear fine-tuned with respect to  seemingly minor details of the nucleon-nucleon interaction, in particular
to the locality of the nucleon-nucleon force~\cite{Elhatisari:2016owd}. By a local force, we refer to the case where
the positions of the particles remain unchanged by the interaction process, while a non-local force
in general induces changes in the particle positions. Our
first objective is to study to what extent the spectrum of $^{12}$C can be described 
by a Wigner SU(4) symmetric two-nucleon force with local as well as non-local components. The strengths
of these two interaction components are tuned to  correctly reproduce the $^{4}$He and $^{12}$C ground-state energies.

The observed cluster phenomena in nuclear systems such as $^{12}$C, see e.g.~\cite{Gai:2014xra},
and the effective interactions between $\alpha$ clusters in nuclei are closely related to the locality
of the  Wigner SU(4) symmetric nucleon-nucleon interaction. For a review on clustering effects in light
nuclei, see Ref.~\cite{Freer:2017gip} and recent related experimental findings are given in
Refs.~\cite{Smith:2020nkz,Li:2020zsd}. The ground state of $^{8}$Be is a  quasi-bound $2\alpha$ state formed by
a short-range repulsion and a medium-range attraction of the effective $\alpha$-$\alpha$ interaction, which
has been experimentally determined from the $\alpha\alpha$ scattering phase shifts. This $\alpha$-$\alpha$
interaction also describes the $3\alpha$ structure of the  ``Hoyle state'', \textit{i.e.} the $0^+_2$ state of $^{12}$C,
as determined by previous NLEFT studies~\cite{Epelbaum:2011md,Epelbaum:2012qn}.
Our second objective is to study to what extent an (appropriately tuned)
Wigner SU(4) symmetric nucleon-nucleon interaction with local and non-local components can describe the structure
and excitation energies of the low-lying states of $^{12}$C, filling a gap of earlier studies that exclusively
considered the even-parity states in terms of 3$\alpha$ clusters.
To this end, we employ PMC simulations with $\alpha$ cluster and harmonic oscillator (HO) trial states,
projected onto different irreducible representations (or \textit{irreps}) of 
the cubic group. This allows us to draw conclusions about the spin-parity quantum numbers $J^P$ and the
(dominant) structure of the states in question. In particular, we are
able to verify the earlier conclusions from NLEFT calculations about the $\alpha$ cluster structure
of the low-lying $0^+$ and $2^+$ states of $^{12}$C. As noted above, we use a simplified SU(4)-symmetric interaction of the form defined in Ref.~\cite{Frame:2020mvv} with local and non-local two-nucleon interactions.  The results we present here can be used as a benchmark for {\rm ab initio} chiral effective field theory calculations on the lattice starting from an accurate description of nucleon-nucleon scattering \cite{Li:2018ymw}.

This paper is organized as follows. In Sec.~\ref{sec:int}, we briefly review the interaction underlying
this study. Section~\ref{sec:methods} contains the details of the calculation of the various excited states
in $^{12}$C. In Sec.~\ref{sec:res} we display the pertinent results for all level up to $\sim 15\,$MeV above the ground state.
We also discuss the interplay between $\alpha$ cluster and shell-model states. These results are
discussed and put into perspective in Sec.~\ref{sec:disc}.


\section{Interactions \label{sec:int}}

We have considered two choices of the spatial lattice spacing, a coarse one of $a = 1.97$~fm with temporal lattice spacing $a_t = 0.66$~fm,
and a finer one of $a = 1.64$~fm with temporal lattice spacing $a_t = 0.55$~fm. The spatial lattice spacings correspond to momentum cut-offs of 314~MeV and
378~MeV, respectively. All calculations use a periodic box with length $L = 9$ in each spatial dimension.
The physical box size for $a = 1.97$~fm is then $17.7$~fm, and for $a = 1.64$~fm it is $14.8$~fm.
This is sufficiently large to suppress finite volume effects to a level that is smaller than the other sources of error in our analysis.

We use an $NN$ interaction $V$ with Wigner SU(4) symmetry \cite{Wigner:1937},
\begin{equation}\label{eq:vsu4}
  V = \frac{C_0}{2} \sum_{\mathbf{n}',\mathbf{n},\mathbf{n}''}
  :\rho_{\rm NL}(\mathbf{n}') f_{s_{\rm L}}(\mathbf{n}'-\mathbf{n})
  f_{s_{\rm L}}(\mathbf{n}-\mathbf{n}'') \rho_{\rm NL}(\mathbf{n}''):,
\end{equation}
where $C_0$ is a coupling constant, vector $\mathbf{n}$ labels the lattice sites, and the colons denote normal ordering. The function $f_{s_{\rm L}}$ is defined with a
``local smearing'' parameter $s_{\rm L}$ as
\begin{subnumcases}
{f_{s_{\rm L}}(\mathbf{n}) =} 1, & $|\mathbf{n}| = 0$, \\
s_{\rm L}, & $|\mathbf{n}| = 1$, \\
0, & otherwise.
\end{subnumcases}
The non-local density operator $\rho_{\rm NL}(\mathbf{n})$ in Eq.~(\ref{eq:vsu4}) is
\begin{equation}\label{eq:NLdens}
  \rho_{\rm NL}(\mathbf{n}) = a_{\rm NL}^\dagger(\mathbf{n}) a_{\rm NL}^{}(\mathbf{n}),
\end{equation}
with the non-local creation and annihilation operators defined with the ``non-local smearing''
parameter $s_{\rm NL}$ as
\begin{align}
  a_{\rm NL}^\dagger(\mathbf{n}) &= a^\dagger(\mathbf{n})
  + s_{\rm NL} \sum_{|\mathbf{n}'|=1} a^\dagger(\mathbf{n}+\mathbf{n}'), \\
  a_{\rm NL}(\mathbf{n}) &= a(\mathbf{n})
  + s_{\rm NL} \sum_{|\mathbf{n}'|=1} a(\mathbf{n}+\mathbf{n}').
\label{eq:ops}
\end{align}

We note that in this highly simplified nuclear interaction certain components are missing, which should be included in a more realistic calculation. While we do not consider a
three-nucleon force, we find that its effects can mostly be absorbed into a renormalization of the strength of the two-nucleon force. This is consistent with our finding that the ratio of $\langle :\rho^3: \rangle $ to $\langle :\rho^2: \rangle $ is approximately the same for all nuclear states of $^{12}$C. Similarly, the Coulomb interaction does not contribute much to the energy splitting between $^{12}$C levels, and so its contribution to the binding energy can also be absorbed by a renormalization of the two-nucleon force.

Specifically, the two interaction parameters $C_0, s_{\rm L}$, are determined by fitting to the ground-state energies of $^{4}$He and $^{12}$C, while $s_{\rm NL} = 0.2$ is taken to equal that used in Ref.~\cite{Frame:2020mvv}. The fitted parameters and ground state energies of $^{4}$He and $^{12}$C are given in Table~\ref{tab:fit}. 

\renewcommand{\arraystretch}{1.2}
\begin{table}[!htbp]
  \caption{Fit results of the parameters $C_0$ and $s_{\rm L}$ for lattice spacings $a = 1.97$~fm and $1.64$~fm.
  The obtained ground state energies of $^{4}$He and $^{12}$C are also given, in comparison with experiment.}
  \label{tab:fit}
  \centering
  \begin{tabular}{|l|cc|c|}
  \hline
  & $a = 1.97$ fm & $a = 1.64$ fm & Exp. \\
  \hline
  $C_0$ [MeV$^{-2}$] & $-5.53 \times 10^{-6}$ & $-3.72 \times 10^{-6}$ & \\
  $s_{\rm L}$   & $0.073$ & $0.083$ & \\
  $E_{^{4}\text{He}}$ [MeV] & $-28.299 (9) $  & $-28.290 (15)$ & $-28.296 $ \\
  $E_{^{12}\text{C}}$ [MeV] & $-92.15 (3) $ & $-92.12 (4)$ & $-92.162 $ \\
  \hline
  \end{tabular}
\end{table}


\section{Methods \label{sec:methods}}

Let us first discuss the basics of our PMC simulations. We define the
transfer matrix operator
\begin{equation}\label{eq:tM}
  M = :\exp(-\alpha_t H):,
\end{equation}
with $\alpha_t = a_t/a$ the ratio of temporal to spatial lattice spacings. The Hamiltonian is given by
\begin{equation}\label{eq:H}
  H = T + V,
\end{equation}
where the interaction $V$ corresponds to Eq.~(\ref{eq:vsu4}), and the kinetic energy $T$ is taken to be
\begin{align}
  T =& \frac{3}{m_N} \sum_{\mathbf{n}} \rho(\mathbf{n}) \notag \\
  & - \frac{1}{2m_N} \sum_{\mathbf{n}} \sum_{l=1}^3
  \left[ \rho(\mathbf{n},\mathbf{n}+e_l) + \rho(\mathbf{n},\mathbf{n}-e_l) \right],
\label{eq:T}
\end{align}
with $e_l$ a unit vector along the $l$-axis, and $m_N = 938.92$~MeV is the nucleon mass. 
The density operator is
\begin{align}
  \rho(\mathbf{n}) &= a^\dagger(\mathbf{n}) a(\mathbf{n}), \\
  \rho(\mathbf{n},\mathbf{n}') &= a^\dagger(\mathbf{n}) a(\mathbf{n}').
\label{eq:densOP}
\end{align}

For a PMC simulation with $N_{\rm ch}$ coupled ``channels'', we start from a set of initial trial states $|\Phi_i\rangle$ with $i = 1, 2,
\dots, N_{\rm ch}$. One can define the Euclidean projection amplitudes at time step $N_t$ as
\begin{equation}\label{eq:Zkl}
  Z_{kl}(N_t) = \langle \Phi_k| M^{N_t} |\Phi_l \rangle.
\end{equation}
Each of the trial states $|\Phi_i\rangle$ is a Slater determinant of single-particle orbitals.
These should be chosen properly, in order to get good statistics for the eigenstates of $H$ at small
enough projection time, before any sign problem becomes severe.

By means of the projection amplitudes (\ref{eq:Zkl}), we construct the ``adiabatic'' transfer matrix
\begin{equation}\label{eq:M}
  M_{qq'}^{(a)}(N_t) = \sum_{q''} Z_{qq''}^{-1}(N_t) Z_{q''q'}^{}(N_t+1),
\end{equation}
with eigenvalues
\begin{equation}\label{eq:EV}
  \lambda_i(N_t) = \exp(-\alpha_t E_i(N_t)),
\end{equation}
such that the low-energy spectrum is given by the ``transient energies''
\begin{equation}\label{eq:energies}
  E_i(N_t) = - \frac{\log(\lambda_i(N_t))}{\alpha_t},
\end{equation}
in the limit of large projection time.
To extrapolate the transient energies to $t = N_t a_t \to \infty$, the following \textit{ansatz} (similar to Ref.~\cite{Lahde:2014sla}) is used:
\begin{equation}\label{eq:extrap}
  E_i(t) = \frac{E_i+\displaystyle\sum_{k=1}^{k_{\rm max}} (E_i+\Delta E_k) c_{i,k}e^{-\Delta E_k t}}{1+\displaystyle\sum_{k=1}^{k_{\rm max}} c_{i,k}e^{-\Delta E_k t}},
\end{equation}
where $E_i, \Delta E_k, c_{i,k}$ are fit parameters.
The choice of $k_{\rm max}$ depends on the details of the fitting procedure.

On the lattice, the full rotational symmetry group is reduced to the finite cubic group, and the $2J+1$
multiplet for states with angular momentum $J$ will split into subgroups of different irreducible
representations (\textit{irreps}).
The decompositions of the first few $(J\leq 3)$ \textit{irreps} with spherical harmonics $Y_{l,m}$ are
\cite{Johnson:1982yq,Lu:2014xfa}:
\begin{subequations}\label{eq:J0123}\begin{align}
  J = 0&:~ A_1\left[ Y_{0,0} \right], \label{eq:J0} \\
  J = 1&:~ T_1\left[Y_{1,0}, Y_{1,\pm 1}\right], \label{eq:J1} \\
  J = 2&:~ E\left[\sqrt{\frac{1}{2}}Y_{2,2}+\sqrt{\frac{1}{2}}Y_{2,-2}, Y_{2,0}\right] \notag \\
      &~~~ \otimes T_2\left[\sqrt{\frac{1}{2}}Y_{2,2}-\sqrt{\frac{1}{2}}Y_{2,-2}, Y_{2,\pm 1}\right], \label{eq:J2} \\
  J = 3&:~ A_2\left[\sqrt{\frac{1}{2}}Y_{3,2}-\sqrt{\frac{1}{2}}Y_{3,-2}\right] \notag \\
      &~~~ \otimes T_1\left[\sqrt{\frac{5}{8}}Y_{3,\mp 3}+\sqrt{\frac{3}{8}}Y_{3,\pm 1}, Y_{3,0}\right] \notag \\
  &~~~ \otimes T_2\left[\sqrt{\frac{5}{8}}Y_{3,\pm 1}-\sqrt{\frac{3}{8}}Y_{3,\mp 3}, \sqrt{\frac{1}{2}}Y_{3,2}
    +\sqrt{\frac{1}{2}}Y_{3,-2}\right] \label{eq:J3}.
\end{align}\end{subequations}
When the initial state $|\Phi_i\rangle$ does not possess a good angular momentum $J$ and projection (along $z$-axis) $J_z$, we project it explicitly onto a specific \textit{irrep}.

Two types of initial states will be considered here. The first ones are based on $\alpha$~clusters with spatially distributed Gaussian wave packets: 
\begin{equation}\label{eq:}
  \phi(\mathbf{r}) = \exp\left(-\frac{\mathbf{r}^2}{2w^2}\right),
\end{equation}
with $w$ the width of the wave packet. The second ones
are shell-model states with HO wave functions. The specific choice of each basis will be discussed in detail below. This is new compared to earlier
NLEFT studies of the $^{12}$C spectrum, which focused entirely on even-parity $\alpha$~cluster states.
Because of this extended basis, we will be able to investigate all excited levels up to a given excitation
energy (about $15\,$MeV).

\begin{figure}[!t]
  \centering
  \includegraphics[width=3.8cm]{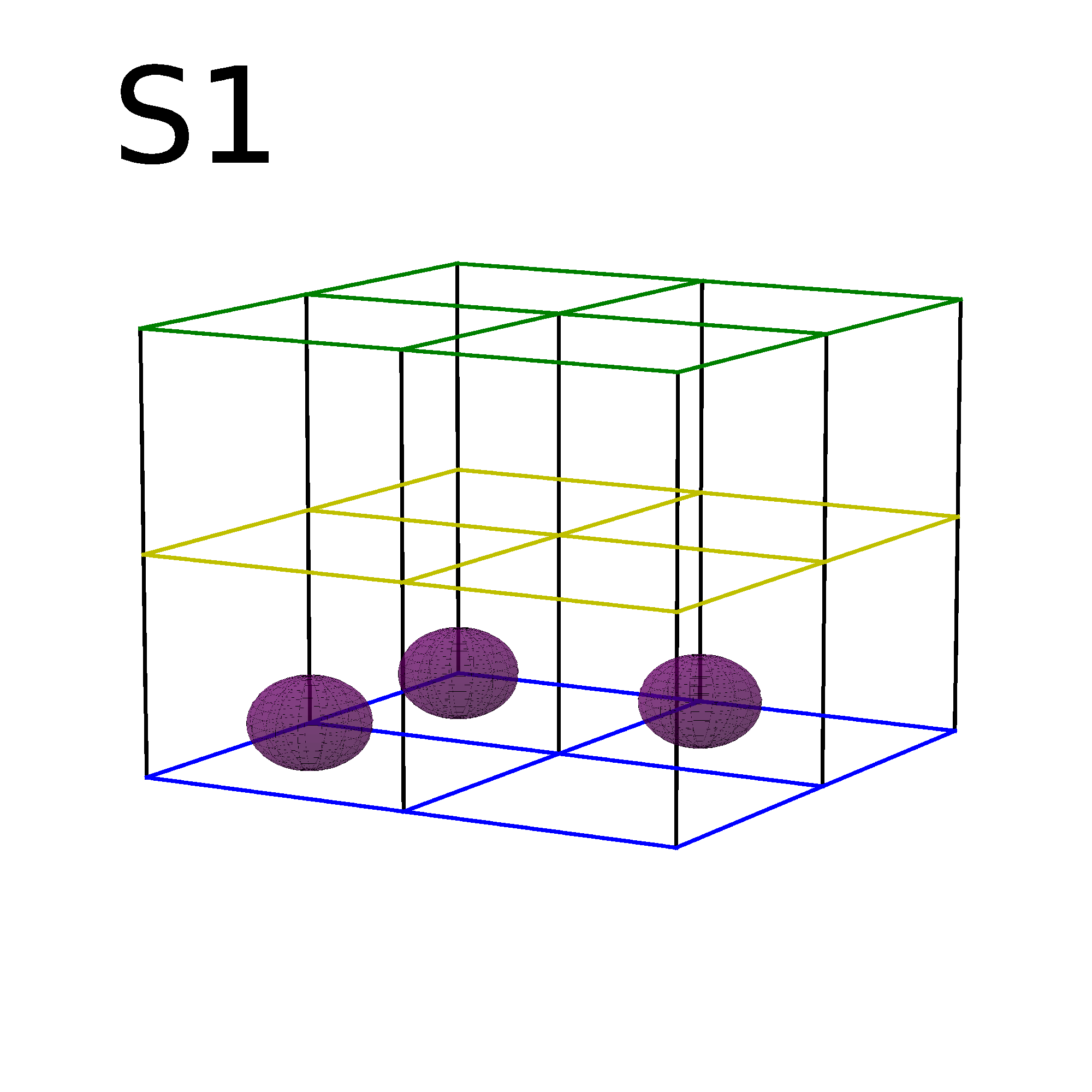}
  \includegraphics[width=3.8cm]{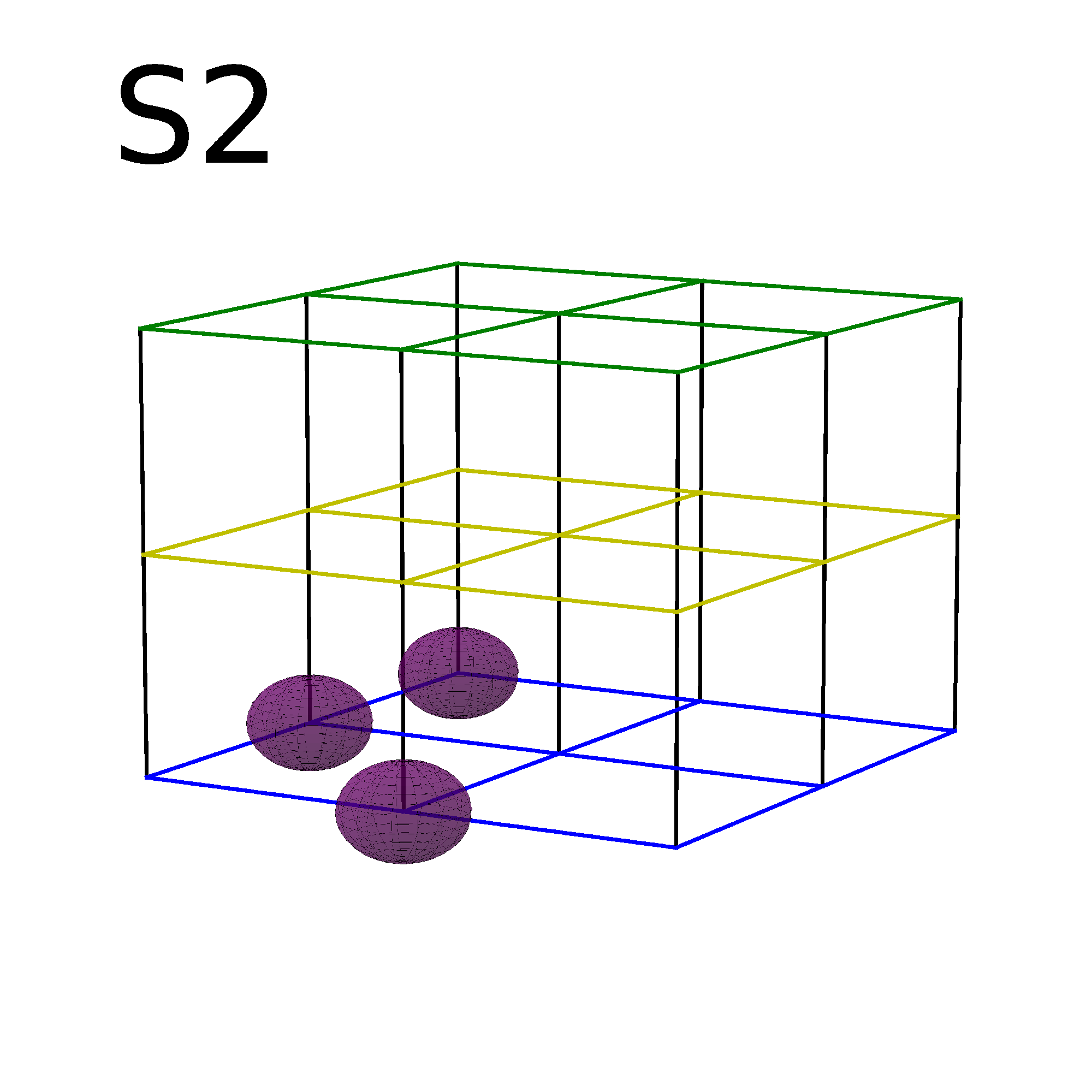}
  \includegraphics[width=3.8cm]{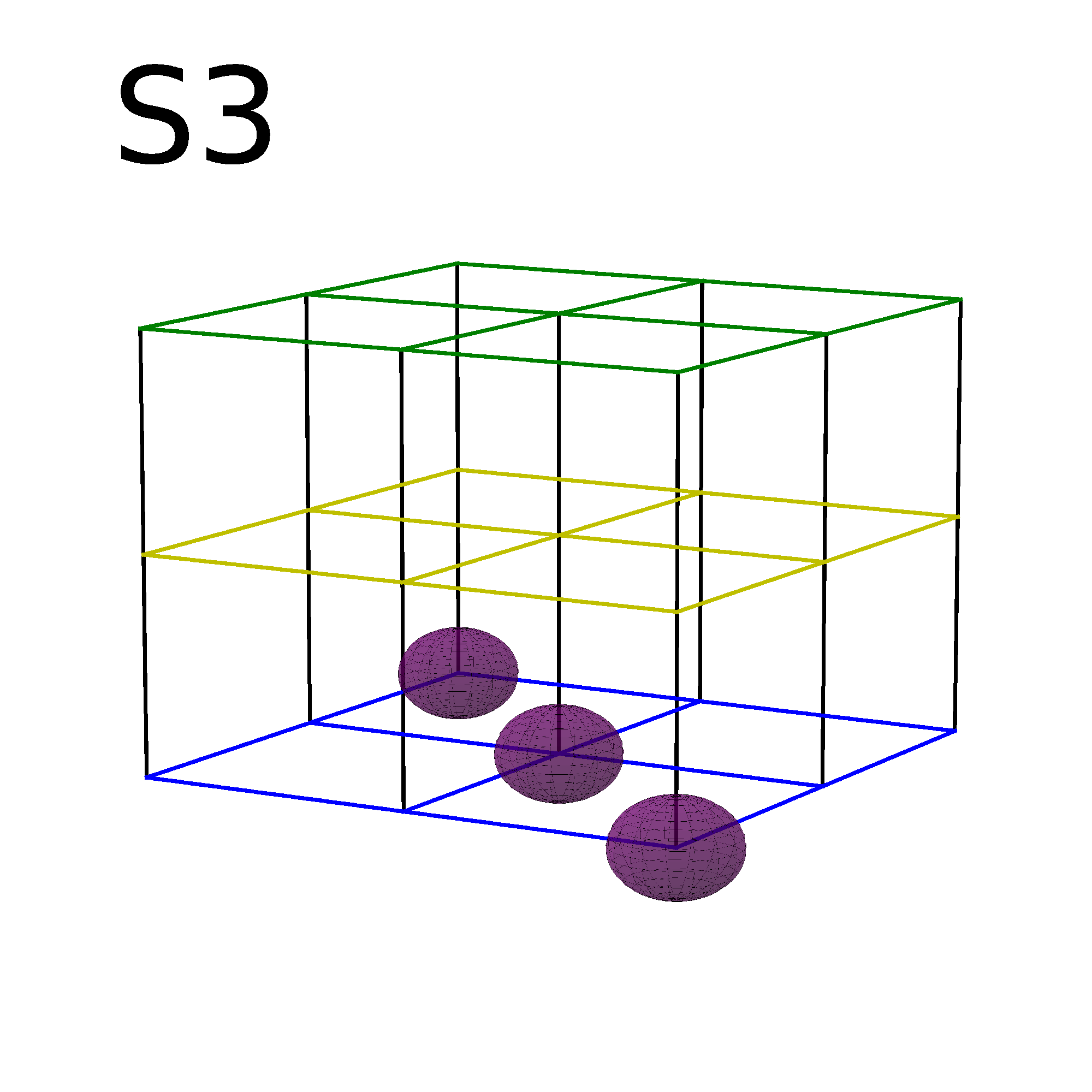}
  \includegraphics[width=3.8cm]{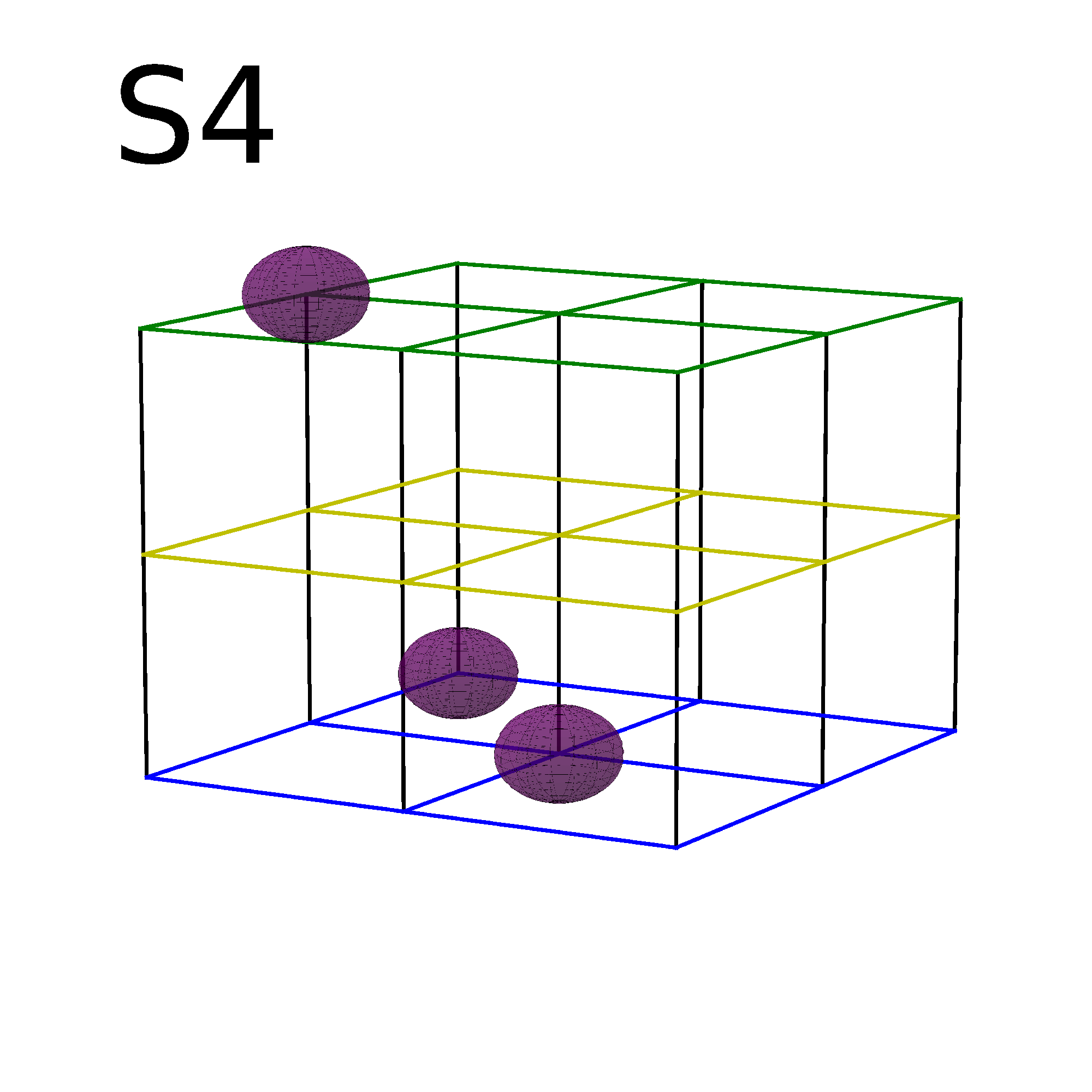}
  \caption{Spatial configurations for 3-$\alpha$ cluster states S1, S2, S3, and S4.
  For S1, S2, and S3, all the three $\alpha$ clusters are on the bottom layer, while for S4 $1 \alpha$ is located on the top layer and $2 \alpha$ on the bottom layer.
  S1 is a isosceles right triangle, S2 is of ``bent-arm'' shape, S3 is a linear chain along the diagonal direction, and S4 is an acute isosceles triangle.}
  \label{fig:s1-s4}
\end{figure}

The 3-$\alpha$ cluster states used in this work are shown in Fig.~\ref{fig:s1-s4}.
These correspond to isosceles right triangles (S1), ``bent-arm'' (obtuse triangular) configurations (S2),
linear chains (S3), and acute isosceles triangles (S4). We note that linear chain configurations have
also been used in DFT studies of $\alpha$~cluster states, see Ref.~\cite{Ren:2020prd}.


\section{Results \label{sec:res}}

In this section, we show the results for various {\it irreps} that allow us to extract a number of excited states, using alpha-cluster and shell-model basis states \cite{Lu:2021}. In case of the shell model states, such a projection can be avoided
by constructing initial states with definite $J_z$. This is in particular useful for the extraction of the higher spin
states, as shown below.

\subsection{The \textit{irrep} $A_1^+$}

\begin{figure}[!htbp]
  \centering
  \includegraphics[width=7.0cm]{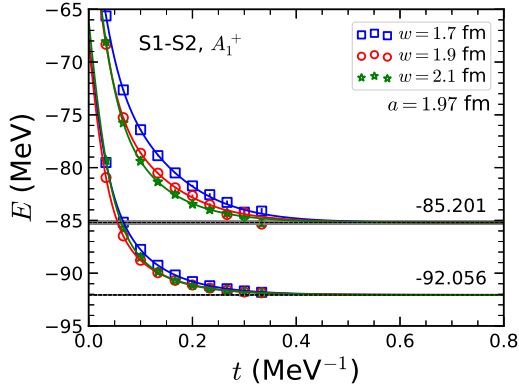}
  \caption{Transient energies of first and second $0^+$ state obtained by two-channel PMC with $A_1^+$ projection
    (open symbols) versus Euclidean projection time.
    Three-$\alpha$ cluster configurations labelled as S1 and S2 are used as initial states and their spatial
    configurations are shown in Fig.~\ref{fig:s1-s4}.
    Three sets of data are calculated using different wave-packet width $w$, and extrapolation fits
    using Eq.~(\ref{eq:extrap}) are given by corresponding lines.
     The error of the extrapolated value is indicated by the gray band.}
  \label{fig:s1-s2}
\end{figure}

First, we show the results of a two-channel PMC simulation with $A_1^+$ projection (corresponding to the
two lowest $0^+$ states), using the
3-$\alpha$ cluster configurations S1 and S2 (see Fig.~\ref{fig:s1-s4}) as initial states.
The obtained transient energies versus Euclidean projection time are shown in Fig.~\ref{fig:s1-s2}.
Three sets of data are obtained using wave-packet widths of $w = 1.7$, $1.9$, and $2.1$~fm, and the extrapolation fits using Eq.~(\ref{eq:extrap}) are given by the corresponding lines.
The number of exponentials in Eq.~(\ref{eq:extrap}) in this case is chosen as $k_{\rm max} = 3$, and will be the same for the other cases discussed in this paper unless otherwise stated.
The error of the extrapolated value is indicated by the horizontal gray band.

The first $0^+$ state (ground state) and second $0^+$ state (Hoyle state) can be identified.
As has been found in previous works \cite{Epelbaum:2011md,Epelbaum:2012qn}, the ``bent-arm'' structure of S2
is suitable to search for the Hoyle state.
This has been confirmed with current simulation with \textit{irrep} projection of $A_1^+$.

\begin{figure}[!htbp]
  \centering
  \includegraphics[width=3.75cm]{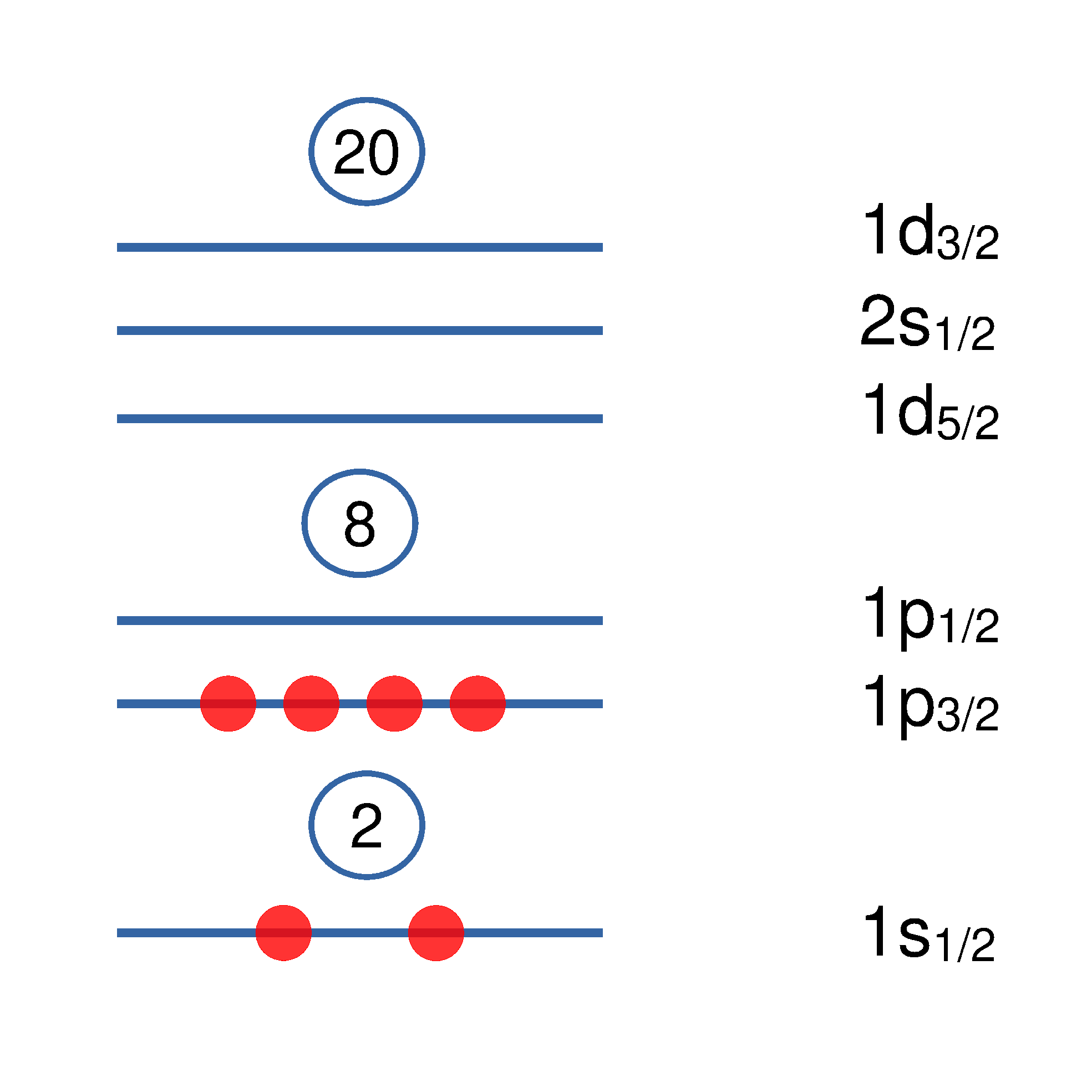} \hspace{1cm}
  \includegraphics[width=3.75cm]{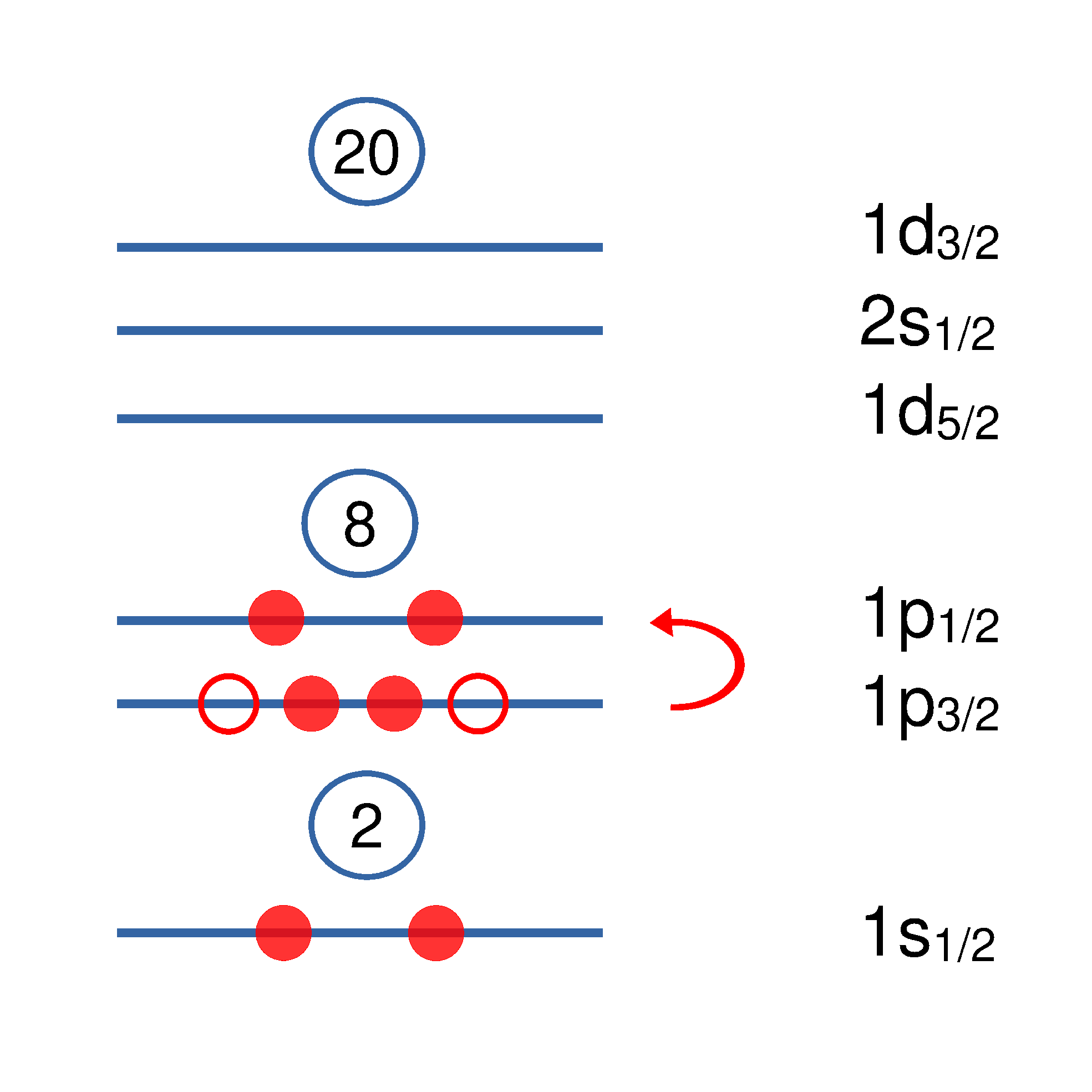}
  \caption{Schematic drawing of $^{12}$C proton shell-model trial states:
    (Left) ground state $|0\rangle$; (Right)
    2p-2h state $|2p_{1/2}^{}2p_{3/2}^{-1}\rangle$.}
  \label{fig:h1-h2}
\end{figure}

\begin{figure}[!htbp]
  \centering
  \includegraphics[width=7.0cm]{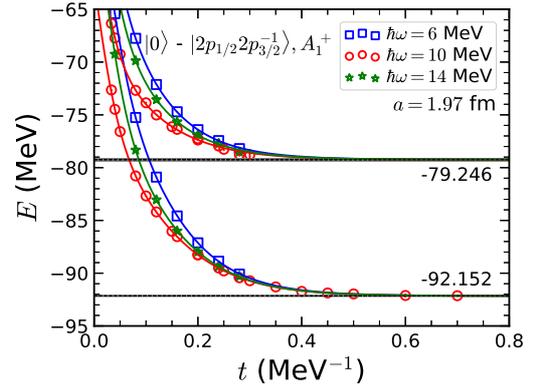}
  \caption{Transient energies of first and third $0^+$ states obtained by two-channel PMC with $A_1^+$ projection
    (open symbols) versus projection time.
  Shell-model wave functions are used as initial states and their configurations are shown in Fig.~\ref{fig:h1-h2}.
  Three sets of data are calculated using HO states with different strength $\hbar\omega$.}
  \label{fig:h1-h2-A1+}
\end{figure}

To show the importance of the choice of initial states, we performed another two-channel PMC simulation with $A_1^+$
projection, using shell model particle-hole ($|ph^{-1}\rangle$) states with HO wave functions.
The configurations of protons for the two channels are shown in Fig.~\ref{fig:h1-h2}.
The configurations of neutrons are always the one with the lowest energy (no particle-hole excitation) unless
otherwise stated. For the first channel, the ground state $|0\rangle$ is chosen with neutrons/protons fully
occupied from $1s_{1/2}$ to $1p_{3/2}$, while for the second channel a proton 2-particle-2-hole (2p-2h) state
of $|2p_{1/2}^{}2p_{3/2}^{-1}\rangle$ with $J_z = 0$ is used.
The reason for the choice of a 2-particle-2-hole is simply because this is found to give the lowest possible energy
for the excited $0^+$. A 1-particle-1-hole (1p-1h) state of $|p_{1/2}^{}p_{3/2}^{-1}\rangle$ can only couple to $1^+$ and $2^+$. The 1p-1h states of $|s_{1/2}^{}s_{1/2}^{-1}\rangle$ or $|p_{3/2}^{}p_{3/2}^{-1}\rangle$ also give an excited $0^+$ state, but with a much higher energy.
The transient energies versus projection time are shown in Fig.~\ref{fig:h1-h2-A1+}.
Three sets of data are calculated using HO strengths $\hbar\omega = 6, 10$, and $14$~MeV, and extrapolation
fits using Eq.~(\ref{eq:extrap}) are given by the corresponding lines.

In Fig.~\ref{fig:h1-h2-A1+}, we also obtained a strong signal for the ground state, consistent with the one obtained in Fig.~\ref{fig:s1-s2}. However, the excited $0^+$ state now has a much higher energy ($-79.2$~MeV)
than the one obtained with the bent-arm 3-$\alpha$ cluster structure ($-85.2$~MeV).
The second channel of the simulation with the 2p-2h shell-model state does not find the Hoyle state either, but again gives us a $0^+$ state with higher energy. This finding is consistent with observations within the no-core shell model~\cite{Navratil:2000ww,Navratil:2007we,Roth:2011ar,Barrett:2013nh}.

\subsection{The \textit{irrep} $E^+$} \label{sec:ep}

\begin{figure}[!htbp]
  \centering
  \includegraphics[width=7.0cm]{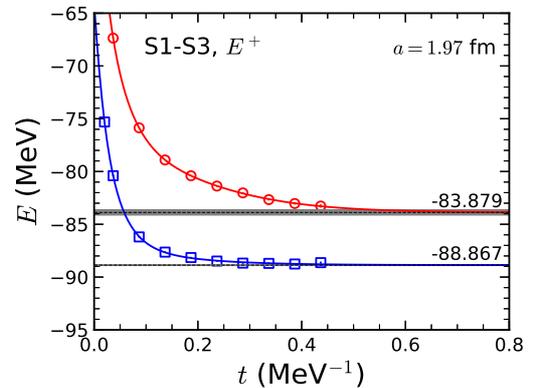}
  \caption{Transient energies of first and second $2^+$ state obtained by two-channel PMC with $E^+$ projection
    (open symbols) versus Euclidean projection time.}
  \label{fig:s1-s3}
\end{figure}

On the one hand, while the cluster states S1 and S2 give strong signals for the first and second $0^+$ states, the signals
for the $2^+$ states with $E^+$ projection with the same cluster states are much less conclusive. On the other hand, the diagonal linear-chain structure S3 is found to work better in this case.
In Fig.~\ref{fig:s1-s3}, we show the energies obtained by PMC with $E^+$ projection, using
the S1 and S3 configurations as initial states.

In Fig.~\ref{fig:s1-s3}, we can identify the first and second $2^+$ states.
Similar to the case of the $0^+$ states, a two-channel PMC simulation using shell-model states
cannot find the second $2^+$ state, but is helpful in finding one with a higher excitation energy.
As the results look similar to Fig.~\ref{fig:h1-h2-A1+}, we do not show them here. We note that our findings reinforce those of Refs.~\cite{Epelbaum:2011md,Epelbaum:2012qn}, in particular concerning the notion that the second $2^+$ state is a rotational excitation of the Hoyle state.

Using the \textit{irrep} projection technique, we can search for positive parity states as well as negative parity ones, and we shall now turn to the discussion of the $3^-$ state.

\subsection{The \textit{irrep} $A_2^-$}

\begin{figure}[!htbp]
  \centering
  \includegraphics[width=7.0cm]{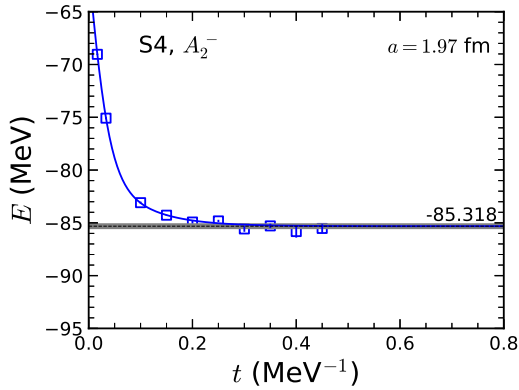}
  \caption{Transient energies of $3^-$ state obtained by PMC with $A_2^-$ projection (open symbols) versus
    projection time. The spatial configuration S4 of the initial state is shown in Fig.~\ref{fig:s1-s4}.}
  \label{fig:s4}
\end{figure}

From Eq.~(\ref{eq:J3}), one finds that $A_2^-$ projection allows for a determination of the $3^-$ state.
The cluster structure S4 is found to be suitable in this calculation, the shape of which is shown in
Fig.~\ref{fig:s1-s4}. Notably, S4 is similar to the
``pear shape'' of the spherical harmonics $Y_{3\pm 2}$.
The simulation results with the S4 trial state are shown in Fig.~\ref{fig:s4}.
As can be seen, the convergence of the energy versus projection time is fast and is achieved
already around $t \simeq 0.2$~MeV$^{-1}$.
The number of exponentials in Eq.~(\ref{eq:extrap}) in this case is  $k_{\rm max} = 2$, due to the fast convergence.

It should be noted that different projections in Eq.~(\ref{eq:J0123}) can be used to obtain other
excited states, given properly chosen initial cluster states. In some cases, shell-model states work better and \textit{irrep} projection is not needed.

\subsection{Without projection}

\begin{figure}[!htbp]
  \centering
  \includegraphics[width=3.95cm]{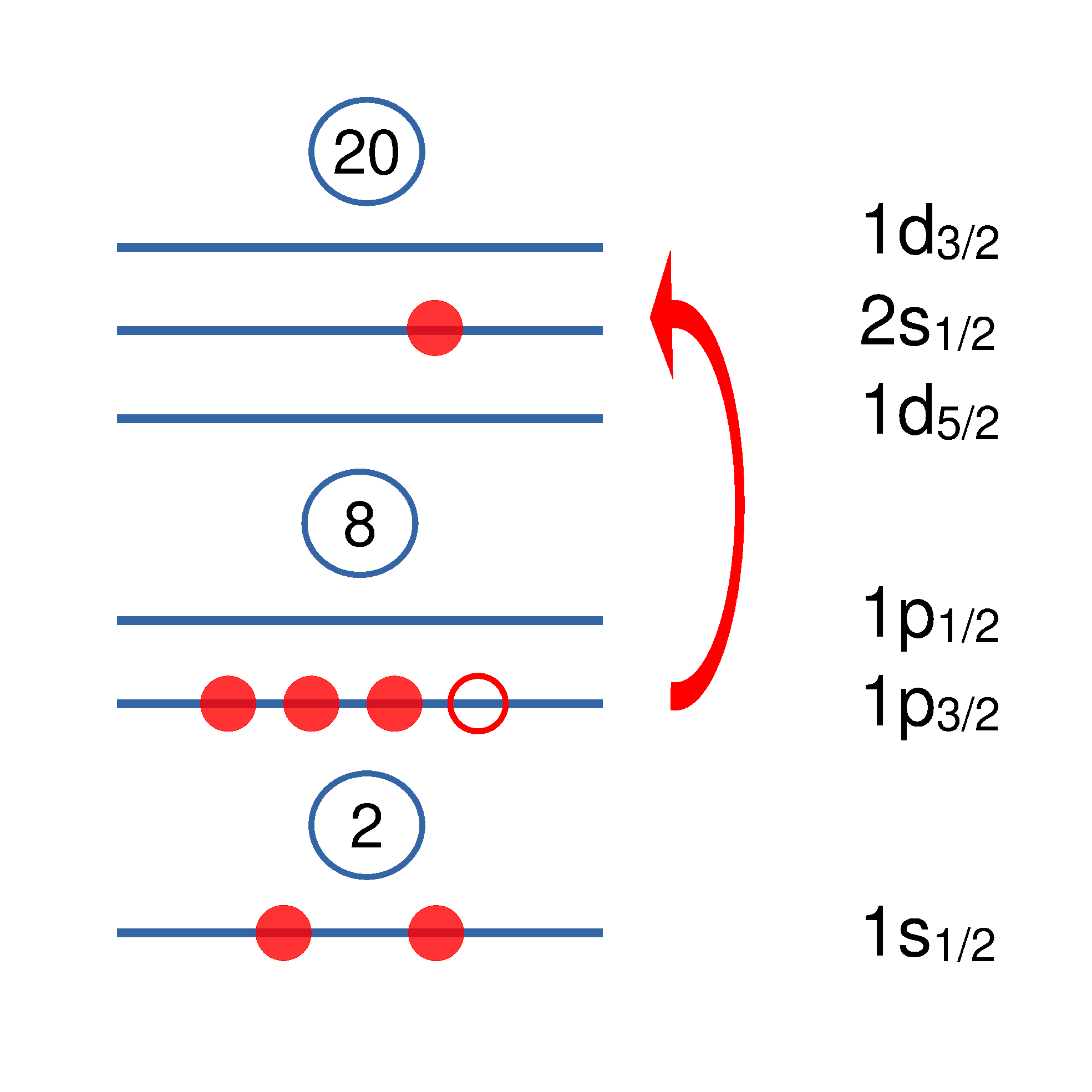}
  \caption{Shell-model 1p-1h configuration $|s_{1/2}^{}p_{3/2}^{-1}\rangle$ of $^{12}$C used to
    search for $1^-$ and $2^-$ states.}
  \label{fig:h3}
\end{figure}

\begin{figure}[!htbp]
  \centering
  \includegraphics[width=7.0cm]{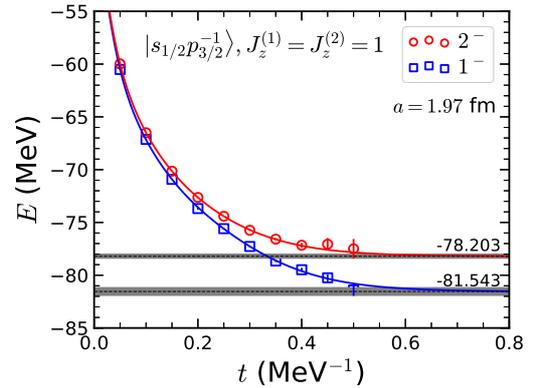}
  \caption{Transient energies of the $1^-$ and $2^-$ states, obtained by two-channel PMC without \textit{irrep} projection (open symbols) versus Euclidean projection time.
    The initial states of the two channels are 1p-1h state $|s_{1/2}^{}p_{3/2}^{-1}\rangle$,
    Fig.~\ref{fig:h3}, both with $J_z = 1$.}
  \label{fig:h3-n}
\end{figure}

When shell-model particle-hole wave functions are used, it is easy to construct the initial state with given
angular momentum projection $J_z$.
In the following example, we construct two states for a two-channel simulation using the 1p-1h
state $|s_{1/2}^{}p_{3/2}^{-1}\rangle$, see Fig.~\ref{fig:h3}, both channels having $J_z = 1$.
It is straightforward to see that this choice should give us the $1^-$ and $2^-$ states. With the choice of the same $J_z$ for both channels, the Euclidean time projection converges to the $1^-$ and $2^-$ states, instead of to two (degenerate) $1^-$ or $2^-$ states.
Note that two states with both $J_z = 0$ will work similarly.
The results are shown in Fig.~\ref{fig:h3-n}.

At small projection times, the energies in the two channels are close to each other, but the splitting eventually increases. At this point, one still cannot distinguish which one is $1^-$ and $2^-$.
A separate one-channel simulation with the same configuration but $J_z = 2$ is sufficient to uniquely identify the $2^-$ state, and the other one will then be the $1^-$ state.
The results of the one-channel simulation with $J_z = 2$ (not shown here) give the same energies as
the red circles in Fig.~\ref{fig:h3-n}, hence definite $J^\pi$ values can be assigned.

Let us summarize how some of the other excited states can be obtained in a similar way:
\begin{itemize}
\item $1^+$: 1p-1h state $|p_{1/2}^{}p_{3/2}^{-1}\rangle$. Two-channel trial states are constructed, both
with $J_z = 0$ (or 1). The PMC simulation gives the $1^+$ and $2^+$ states.
Since the $2^+$ state has been identified as in Sec.~\ref{sec:ep}, the other one will be $1^+$ by default.
\item $4^-$: 1p-1h state $|d_{5/2}^{}p_{3/2}^{-1}\rangle$. A one-channel trial state with $J_z = 4$ is
constructed. Since the largest $J$ value allowed by this 1p-1h configuration is $4$, the PMC
simulation gives the $4^-$ state directly.
\item $4^+$: 2p-2h state $|(\nu\pi)p_{1/2}^{}(\nu\pi)p_{3/2}^{-1}\rangle$.  A one-channel trial state
with $J_z = 4$ is constructed. The reason to choose 2p-2h composed of 1 proton particle-hole
and 1 neutron particle-hole instead of 2 protons (or neutrons) is to assure that the largest $J_z$
value can be $4$. When both are protons (neutrons), the largest $J_z$ value can only be 2.
The PMC simulation gives the $4^+$ state.
\end{itemize}

\subsection{The $^{12}$C spectrum}

\begin{figure*}[t]
  \centering
  \includegraphics[width=.99\textwidth]{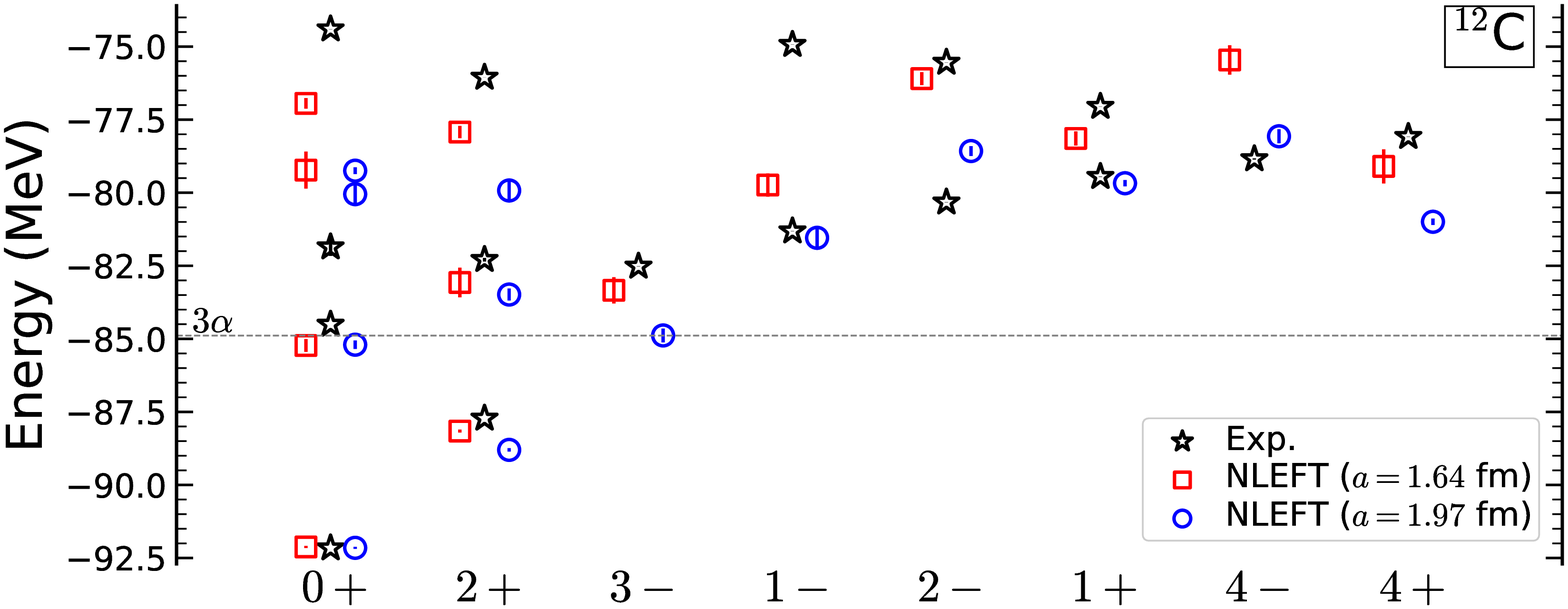}
  \caption{Spectrum of $^{12}$C below $\sim 15$~MeV excitation energy obtained by NLEFT using an SU(4) symmetric $NN$ interaction in comparison with experiment~\cite{Kelley:2017qgh}. Note that the ground state is used for tuning of the interaction.}
  \label{fig:lev}
\end{figure*}

To summarise the above results, the spectrum of $^{12}$C below $\sim 15$ MeV obtained by PMC using SU(4) interaction is shown in Tab.~\ref{tab:lev} and Fig.~\ref{fig:lev}, together with the experimental data from Ref.~\cite{Kelley:2017qgh}.
For $2^+$ and $3^-$, the multiplet-averaged energy is used as in Ref.~\cite{Lu:2014xfa}.
It is quite interesting to see that with such a simple SU(4) symmetric interaction, which was fitted to the ground states of $^{4}$He and $^{12}$C, all the levels of $^{12}$C below 15~MeV have been found, and the agreement with experiment is surprisingly good. This suggests that our SU(4) symmetric interaction successfully incorporates many key features of the $^{12}$C system.

\begin{table}[!htbp]
  \caption{Energies of the low-lying states in $^{12}$C calculated by NLEFT using an SU(4) symmetric $NN$ interaction.
  Results for two lattice spacings are shown, in comparison with experiment~\cite{Kelley:2017qgh}.
  Energies are given in MeV.
  Errors smaller than 1~keV are not displayed.}
  \label{tab:lev}
  \centering
  \begin{tabular}{|c|cc|c|}
  \hline
  State & $a = 1.97$~fm & $a = 1.64$~fm & Experiment \\
  \hline
  $0_1^+$ & $-92.15 (3) $ & $-92.12 (4) $ & $-92.162 $ \\
  $2_1^+$ & $-88.87 (4) $ & $-88.19 (17)$ & $-87.722 $ \\
  $0_2^+$ & $-85.20 (15)$ & $-85.23 (22)$ & $-84.508 $ \\
  $3_1^-$ & $-84.9 (2)  $ & $-83.3 (5)  $ & $-82.521 (5)$ \\
  $2_2^+$ & $-83.5 (2)  $ & $-83.1 (5)  $ & $-82.29 (6)$ \\
  $0_3^+$ & $-80.0 (3)  $ & $-79.2 (6)  $ & $-81.9 (3)$ \\
  $1_1^-$ & $-81.5 (4)  $ & $-79.7 (4)  $ & $-81.315 (4)$ \\
  $2_1^-$ & $-78.6 (2)  $ & $-76.1 (2)  $ & $-80.326 (4)$ \\
  $1_1^+$ & $-79.67 (11)$ & $-78.14 (24)$ & $-79.452 (6)$ \\
  $4_1^-$ & $-78.1 (2)  $ & $-75.5 (5)  $ & $-78.846 (20)$ \\
  $4_1^+$ & $-80.99 (11)$ & $-79.1 (6)  $ & $-78.083 (5)$ \\
  $2_3^+$ & $-79.9 (4)  $ & $-77.9 (2)  $ & $-76.056 $ \\
  $0_4^+$ & $-79.25 (11)$ & $-76.94 (18)$ & $-74.402 $ \\
  \hline
  \end{tabular}
\end{table}


When the lattice spacing $a = 1.97$~fm is decreased to $1.64$~fm, the excitation energies generally increase, with a few exceptions such as the Hoyle state.
The agreement with experimental data for $a = 1.64$~fm is improved for states such as
$2^+$, $3^-$, $4^+$, and $0_4^+$, while for other cases the agreement appears to worsen slightly.
Overall, the spectrum obtained by NLEFT is in quite good agreement with experimental data, given that only a
simple SU(4) symmetric interaction has been used.
Of course, the other components of the nuclear force not included here are very important and crucial to give
a good universal description of the whole nuclear chart.
The chiral EFT interaction up to N3LO for lattice simulation is a work in progress, and will be used in the future to
investigate the spectrum of nuclei using the methods established in this work.


\section{Summary and outlook \label{sec:disc}}

We have explored the low-lying spectrum of $^{12}$C using a simple SU(4)-symmetric interaction with local and non-local terms. By fitting the strength of the interaction and
the local smearing parameter to the ground state energies of $^4$He and $^{12}$C, we have
obtained a good representation of the spectrum up to excitation energies of about
15~MeV. This was achieved using initial states composed of three $\alpha$ clusters, as well as of shell model orbitals.  In particular, we were able to confirm earlier NLEFT results concerning the structure
of the Hoyle state and the second $2^+$ state~\cite{Epelbaum:2012qn}. 
For the Hoyle state, prolate $\alpha$ cluster configurations are very important~\cite{Dreyfuss:2016ezg}, and the second $2^+$ state is consistent with the interpretation as a rotational excitation of the Hoyle state.  Our results provide confirmation that $^{12}$C sits at a fascinating balance point where the competition between the shell structure and clustering produces a low-energy spectrum with qualitatively different types of nuclear states.

The success of these simple interactions in describing all of the low-lying states of $^{12}$C suggests that the tendency towards $\alpha$ clustering is probably not a simple binary attribute that effects some states of $^{12}$C and not others.  It is clear that the effects of $\alpha$ clustering are very prominent for the Hoyle state and the second $2^+$ state, to the extent that their overlap with shell model initial states are so small that they cannot be detected in the lattice Monte Carlo calculations. However, it also appears that spin-orbit interactions are not playing a decisive role for the other $^{12}$C states with good overlap with shell model initial states.  This implies that either spin-orbit interactions are somewhat weak in the $^{12}$C system, or the effects of $\alpha$ clustering are diminishing their influence. This is in agreement with previous {\it ab initio} shell model calculations \cite{Hayes:2003ni,Johnson:2014xda}. To clarify the underlying physics further, it would be very interesting to perform similar studies for $^{16}$O and $^{20}$Ne to see whether an SU(4)-symmetric interaction can fully describe the low-energy spectra of these nuclei.


\section*{Acknowledgments}
\sloppypar{
We thank Serdar Elhatisari, Dillon Frame, Calvin Johnson, Bing-Nan Lu and Gautam Rupak for helpful discussions.
This work was supported by DFG and NSFC through funds provided to the
Sino-German CRC 110 ``Symmetries and the Emergence of Structure in QCD" (NSFC
Grant No.~11621131001, DFG Grant No.~TRR110).
The work of UGM was supported in part by VolkswagenStiftung (Grant no. 93562)
and by the CAS President's International
Fellowship Initiative (PIFI) (Grant No.~2018DM0034).
The work of DL is supported by the U.S. Department of Energy (Grant 
No. DE-SC0013365) and the Nuclear Computational Low-Energy Initiative (NUCLEI) SciDAC project, with computing support from the OLCF through the INCITE award ``Ab-initio nuclear structure and nuclear reactions''.
The authors gratefully acknowledge the Gauss Centre for Supercomputing e.V. (www.gauss-centre.eu) 
for funding this project by providing computing time on the GCS Supercomputer JUWELS 
at J\"ulich Supercomputing Centre (JSC).}



\begin{thebibliography}{99}

\bibitem{Lu:2018bat}
B.~N.~Lu, N.~Li, S.~Elhatisari, D.~Lee, E.~Epelbaum and U.-G.~Mei{\ss}ner,
Phys. Lett. B \textbf{797} (2019), 134863
[arXiv:1812.10928 [nucl-th]].

\bibitem{Wigner:1937}
E.~Wigner,
Phys. Rev. \textbf{51}, 106 (1937)

\bibitem{Elhatisari:2016owd}
S.~Elhatisari, N.~Li, A.~Rokash, J.~M.~Alarc\'on, D.~Du, N.~Klein, B.~N.~Lu, U.-G.~Mei{\ss}ner,
 E.~Epelbaum and H.~Krebs, \textit{et al.}
Phys. Rev. Lett. \textbf{117} (2016) no.13, 132501
[arXiv:1602.04539 [nucl-th]].

\bibitem{Contessi:2017rww}
L.~Contessi, A.~Lovato, F.~Pederiva, A.~Roggero, J.~Kirscher and U.~van Kolck,
Phys. Lett. B \textbf{772}, 839-848 (2017)
[arXiv:1701.06516 [nucl-th]].

\bibitem{Rokash:2016tqh}
A.~Rokash, E.~Epelbaum, H.~Krebs and D.~Lee,
Phys. Rev. Lett. \textbf{118}, no.23, 232502 (2017)
[arXiv:1612.08004 [nucl-th]].

\bibitem{Kanada-Enyo:2020zzf}
Y.~Kanada-En'yo and D.~Lee,
Phys. Rev. C \textbf{103}, no.2, 024318 (2021)
[arXiv:2008.01867 [nucl-th]].

\bibitem{Hayes:2003ni}
A.~C.~Hayes, P.~Navratil and J.~P.~Vary,
Phys. Rev. Lett. \textbf{91}, 012502 (2003)
[arXiv:nucl-th/0305072 [nucl-th]].

\bibitem{Johnson:2014xda}
C.~W.~Johnson,
Phys. Rev. C \textbf{91}, no.3, 034313 (2015)
[arXiv:1409.7355 [nucl-th]].

\bibitem{Freer:2017gip}
M.~Freer, H.~Horiuchi, Y.~Kanada-En'yo, D.~Lee and U.-G.~Mei{\ss}ner,
Rev.\ Mod.\ Phys. \textbf{90} (2018) no.3, 035004.

\bibitem{Frame:2020mvv}
D.~Frame, T.~A.~L\"ahde, D.~Lee and U.-G.~Mei{\ss}ner,
Eur. Phys. J. A \textbf{56} (2020) no.10, 248
[arXiv:2007.06335 [nucl-th]].


\bibitem{Kaplan:1995yg}
D.~B.~Kaplan and M.~J.~Savage,
Phys. Lett. B \textbf{365}, 244-251 (1996)
[arXiv:hep-ph/9509371 [hep-ph]].


\bibitem{Kaplan:1996rk}
D.~B.~Kaplan and A.~V.~Manohar,
Phys. Rev. C \textbf{56}, 76-83 (1997)
[arXiv:nucl-th/9612021 [nucl-th]].


\bibitem{Lee:2020esp}
D.~Lee, S.~Bogner, B.~A.~Brown, S.~Elhatisari, E.~Epelbaum, H.~Hergert, M.~Hjorth-Jensen,
H.~Krebs, N.~Li and B.~N.~Lu, \textit{et al.}
[arXiv:2010.09420 [nucl-th]].

\bibitem{Timoteo:2011tt}
V.~S.~Timoteo, S.~Szpigel and E.~Ruiz Arriola,
Phys. Rev. C \textbf{86} (2012), 034002
[arXiv:1108.1162 [nucl-th]].

%
\bibitem{Lee:2008fa}
D.~Lee,
Prog. Part. Nucl. Phys. \textbf{63} (2009) 117.
[arXiv:0804.3501 [nucl-th]].

\bibitem{Lahde:2019npb}
T.~A.~L\"ahde and U.-G.~Mei{\ss}ner,
Lect. Notes Phys. \textbf{957} (2019), 1.


\bibitem{Gai:2014xra}
M.~Gai, R.~Bijker, M.~Freer, T.~Kokalova, D.~J.~Marin-Lambarri and C.~Wheldon,
J.\ Phys.\ Conf.\ Ser. \textbf{569} (2014) no.1, 012011.

\bibitem{Smith:2020nkz}
R.~Smith, M.~Gai, M.~W.~Ahmed, M.~Freer, H.~O.~U.~Fynbo, D.~Schweitzer and S.~R.~Stern,
Phys.\ Rev.\ C \textbf{101} (2020) no.2, 021302.
[arXiv:2001.07223 [nucl-ex]].

\bibitem{Li:2020zsd}
K.~C.~W.~Li, F.~D.~Smit, P.~Adsley, R.~Neveling, P.~Papka, E.~Nikolskii, J.~W.~Br\"ummer,
L.~M.~Donaldson, M.~Freer and M.~N.~Harakeh, \textit{et al.}
[arXiv:2011.10112 [nucl-ex]].

\bibitem{Epelbaum:2011md}
E.~Epelbaum, H.~Krebs, D.~Lee and U.-G.~Mei{\ss}ner,
Phys. Rev. Lett. \textbf{106} (2011), 192501
[arXiv:1101.2547 [nucl-th]].

\bibitem{Epelbaum:2012qn}
E.~Epelbaum, H.~Krebs, T.~A.~L\"ahde, D.~Lee and U.-G.~Mei{\ss}ner,
Phys. Rev. Lett. \textbf{109} (2012), 252501
[arXiv:1208.1328 [nucl-th]].

\bibitem{Li:2018ymw}
N.~Li, S.~Elhatisari, E.~Epelbaum, D.~Lee, B.~N.~Lu and U.-G.~Mei\ss{}ner,
Phys. Rev. C \textbf{98}, no.4, 044002 (2018)
[arXiv:1806.07994 [nucl-th]].

\bibitem{Lahde:2014sla}
T.~A.~L\"ahde, E.~Epelbaum, H.~Krebs, D.~Lee, U.-G.~Mei{\ss}ner and G.~Rupak,
J. Phys. G \textbf{42} (2015) no.3, 034012
[arXiv:1409.7538 [nucl-th]].


\bibitem{Johnson:1982yq}
R.~C.~Johnson,
Phys. Lett. B \textbf{114} (1982), 147-151.

\bibitem{Lu:2014xfa}
B.~N.~Lu, T.~A.~L\"ahde, D.~Lee and U.-G.~Mei{\ss}ner,
Phys. Rev. D \textbf{90} (2014) no.3, 034507
[arXiv:1403.8056 [nucl-th]].


\bibitem{Ren:2020prd}
Z.~X.~Ren, P.~W.~Zhao and J.~Meng,
Phys. Lett. B \textbf{801} (2020), 135194
[arXiv:2001.02834 [nucl-th]].

\bibitem{Lu:2021}
B.~N.~Lu, {\em private communication} (2021).

%
\bibitem{Navratil:2000ww}
P.~Navr\'atil, J.~P.~Vary and B.~R.~Barrett,
Phys. Rev. Lett. \textbf{84} (2000), 5728-5731
[arXiv:nucl-th/0004058 [nucl-th]].

\bibitem{Navratil:2007we}
P.~Navr\'atil, V.~G.~Gueorguiev, J.~P.~Vary, W.~E.~Ormand and A.~Nogga,
Phys. Rev. Lett. \textbf{99} (2007), 042501
[arXiv:nucl-th/0701038 [nucl-th]].

\bibitem{Roth:2011ar}
R.~Roth, J.~Langhammer, A.~Calci, S.~Binder and P.~Navr\'atil,
Phys. Rev. Lett. \textbf{107} (2011), 072501
[arXiv:1105.3173 [nucl-th]].

\bibitem{Barrett:2013nh}
B.~R.~Barrett, P.~Navr\'atil and J.~P.~Vary,
Prog. Part. Nucl. Phys. \textbf{69} (2013), 131-181.

\bibitem{Kelley:2017qgh}
J.~H.~Kelley, J.~E.~Purcell and C.~G.~Sheu,
Nucl. Phys. A \textbf{968} (2017), 71-253.

\bibitem{Dreyfuss:2016ezg}
A.~C.~Dreyfuss, K.~D.~Launey, T.~Dytrych, J.~P.~Draayer, R.~B.~Baker, C.~M.~Deibel and C.~Bahri,
Phys. Rev. C \textbf{95} (2017) no.4, 044312
[arXiv:1611.00060 [nucl-th]].


\end{thebibliography}
\end{document}